\documentclass[twoside]{ima}
\pdfoutput=1

\usepackage{graphicx}
\usepackage{amsmath,amssymb}
\usepackage{bm}
\usepackage{ifthen}
\usepackage{cite}
\usepackage{subfigure}

\graphicspath{{figs/}}

\newcommand{\COLOR}{true}
\newcommand{\ifcolor}[2]{\ifthenelse{\equal{\COLOR}{true}}{#1}{#2}}

\newcommand{\cee}{\mathbb{C}}

\newcommand{\vel}{\bm{u}}
\newcommand{\zzd}{\mathcal{Z}}

\newcommand{\etal}{\textit{et al.}}

\begin{document}

\title{A LOW-REYNOLDS-NUMBER TREADMILLING SWIMMER NEAR A
  SEMI-INFINITE WALL}
\author{KIORI OBUSE\thanks{Research Institute for Mathematical
    Sciences, Kyoto University, Kyoto, 606-8502, Japan.}
\and JEAN-LUC THIFFEAULT\thanks{Dept.\ of Mathematics,
University of Wisconsin,
Madison, WI 53706, USA.
The work of the second author was supported in part by NSF grant
DMS-0806821.}
}

\maketitle

\begin{abstract}
  We investigate the behavior of a treadmilling microswimmer in a
  two-dimensional unbounded domain with a semi-infinite no-slip
  wall. The wall can also be regarded as a probe or pipette inserted
  into the flow.  We solve the governing evolution equations in an
  analytical form and numerically calculate trajectories of the
  swimmer for several different initial positions and orientations.
  We then compute the probability that the treadmilling organism can
  escape the vicinity of the wall.  We find that many trajectories in
  a `wedge' around the wall are likely to escape.  This suggests that
  inserting a probe or pipette in a suspension of organism may push
  away treadmilling swimmers.
\end{abstract}

\section{Introduction}
\label{Sec:Intro}

The locomotion of microorganisms is an active research area of fluid
dynamics and biology (see for instance the reviews \cite{Lauga2009,
  Pedley1992}). As their motion occurs on very small length scales and
speeds, their dynamics is governed by low-Reynolds-number
hydrodynamics, where inertial forces are negligible in comparison to
the viscous effects of the fluid (Stokes flow).

Many studies deal with such dynamics in unbounded or very large
domains~\cite{Shapere1989, Najafi2004, Avron2005}. In reality,
however, most organisms are in the vicinity of other bodies or
boundaries, where hydrodynamic interactions can have a significant
effect on their motion.  The importance of boundaries has also been
suggested by experimental observations.  For example, some research
suggests that microorganisms are attracted to solid
walls~\cite{Rothschild1963, Winet1984,
  Cosson2003}. Berke~\etal~\cite{Berke2008} measured the steady-state
distribution of \emph{E.~Coli} bacteria swimming between two glass
plates and found a strong increase of the cell concentration at the
boundaries. They also theoretically demonstrated that hydrodynamic
interactions of swimming cells with solid surfaces lead to their
reorientation in the direction parallel to the surfaces.
Lauga~\etal~\cite{Lauga2006} showed that circular trajectories are
natural consequences of force-free and torque-free swimming and
hydrodynamic interactions with the boundary. This leads to a
hydrodynamic trapping of the cells close to the
surface. Drescher~\etal~\cite{Drescher2009} found that when two nearby
Volvox colonies swim close to a solid surface, they attract one
another and can form stable bound states in which they `waltz' or
`minuet' around each other. These observations suggest that, in order
to obtain a comprehensive understanding of low-Reynolds-number
locomotion, it is necessary to study hydrodynamic interactions between
microorganisms and boundaries.

Some of the phenomena stated above have already been confirmed by
numerical simulations~\cite{Ramia1993, HernandezOrtiz2005}. However,
not many physical explanations have been given to the locomotion of
microorganism near boundaries. Berke \etal~\cite{Berke2008} have
captured the swimming microorganisms' attraction to boundaries by
modeling the swimmer as a force dipole singularity. However, contrary
to the experimental findings, the microorganism in this model crashes
into the boundary in finite time.  Or and Murray~\cite{Or2009} studied
the dynamics of low-Reynolds-number swimming organism near a plane
wall.  They analyzed the motion of a swimmer consisting of two
rotating spheres connected by a thin rod as a simple theoretical model
of a `treadmilling' swimming organism.  They found that when the
spheres rotate with different velocities their model has a solution
with steady translation parallel to the wall, and that under small
perturbation the swimmer exhibits a `bouncing' motions parallel to the
wall.  These results have recently been verified experimentally on a
macroscale robotic prototype swimming in a highly viscous
fluid~\cite{Zhang2010}.  Furthermore, Crowdy and Or \cite{Crowdy2010}
have proposed a singularity model for swimming microorganisms placed
near an infinite no-slip boundary. Their model was based on a circular
treadmilling organism which has no means of self-propulsion, that is,
the organism doesn't move unless it interacts with a boundary.  (For
example, the organism may be creating a feeding current.)  They
proposed an appropriate Stokes singularities that represent the flow
field created by this treadmilling organism.  By studying the
interaction between these singularities and the no-slip wall, they
formulated explicit evolution equations for the motion of the
organism, and fully characterized its motion near the wall.  They
found trajectories with a periodic bouncing motion along the wall
which had remarkable similarity to the trajectories shown in Or and
Murray \cite{Or2009}. Crowdy and Samson \cite{Crowdy2010b}, using the
point-singularity model, investigated the dynamics of treadmilling
organism near an infinite no-slip boundary with a gap of a fixed
size. They employed a conformal mapping technique to avoid the
difficulty in treating the image of the treadmilling organism on the
wall. They found that the treadmilling organism can exhibit several
qualitatively different types of trajectories: jumping over the gap,
rebounding from the gap, being trapped near the gap, and escaping the
gap region even when the organism has initial position in the gap.
They also performed a bifurcation analysis in terms of the model
parameters, and demonstrated the presence of stable equilibrium points
in the gap region as well as Hopf bifurcations to periodic bound
states.  This reduced model also exhibited a global gluing bifurcation
in which two symmetric periodic orbits merge at a saddle point into
symmetric bound states having more complex spatio-temporal structure.

In the present paper we examine the dynamics of a treadmilling
organism near a semi-infinite no-slip wall, modeled as a flat plate of
zero thickness.  Though this is a special case of the model of Crowdy
and Samson~\cite{Crowdy2010b}, it deserves separate investigation
because of the simpler equations involved, and because the
semi-infinite wall can be regarded as a probe or pipette inserted in
the system, a common situation in microbiology.  We also analyze the
trajectories in a very different manner to~\cite{Crowdy2010b}, as we
attempt to quantify the probability of escape from the wall's
vicinity, assuming the treadmillers are randomly oriented.

\section{Model of a treadmilling microorganism}
\label{sec:wasModel}

Following previous authors~\cite{Crowdy2010,Crowdy2010b}, we consider
a two-dimensional model for a microorganism in the $(x, y)$-plane,
which we treat as the complex plane with $z \equiv x + i y$.  Our
derivation is a special case of~\cite{Crowdy2010b}, who considered a
microswimmer near a slit or gap in an infinite wall.  Nevertheless, as
mentioned in the introduction, the semi-infinite wall is important
enough to be treated separately, as it arises in the neighborhood of a
probe or a pipette.

The Stokes equations which describe the motion of an incompressible
viscous fluid are
\begin{equation}
  \nabla p = \eta \Delta \vel, \qquad \nabla \cdot \vel = 0,
  \label{eq:Stokes}
\end{equation}
where $\Delta$ is the Laplace operator, $\vel = (u_x, u_y )$ is the
fluid velocity, and $p$ and $\eta$ are the pressure and dynamic
viscosity, respectively. As we are considering a two-dimensional flow,
we can introduce a stream function $\psi$, such that the velocity is
given by~$u_x = \partial \psi /\partial y$, $u_y = - \partial \psi
/\partial x$.  Then the Stokes equations~\eqref{eq:Stokes} reduce to
the biharmonic equation~$\Delta^2 \psi = 0$.  The complex velocity is
$W=u_x + i u_y=- 2i {\partial \psi}/{\partial {\bar z}}$, with
\begin{align}
 W &=  u_x + i u_y  = - 2i \frac{\partial
  \psi}{\partial {\bar z}}
   = f(z) + z \overline{f^\prime(z)} + \overline{g^\prime(z)},
 \label{eq:complexvel}\\
 \psi &= {\rm Im}[{\bar z}f(z) + g(z)].
 \label{eq:sol}
\end{align}
where~$f(z)$ and~$g(z)$ are called Goursat functions; they are
analytic everywhere in the flow domain, except where isolated
singularities are introduced to model swimmers.  For a treadmilling
swimmer, we take $f(z)$ to have a simple pole at $z=z_d$, so that
\begin{equation}
 f(z) = \frac{\mu}{z-z_d} + \ldots,
 \qquad
 g^\prime(z) = \frac{\mu \bar z_d}{(z - z_d)^2}
 + \ldots.
 \label{eq:fg_stresslet_singularpart}
\end{equation}
where $\mu \in \cee$ and the form of~$g^\prime$ is forced by the
requirement that the complex velocity~\eqref{eq:complexvel} have no
higher than a~$|z-z_d|^{-1}$ singularity.  This solution corresponds
to a stresslet of strength $\mu$ at $z_d$.  The ellipses
in~\eqref{eq:fg_stresslet_singularpart} indicate analytic terms.  The
expansion~\eqref{eq:fg_stresslet_singularpart} is the basic solution
for a treadmilling swimmer, which does not have any self-propulsion in
itself, but can move due to its interaction with
boundaries~\cite{Leshansky2007, Or2009, Zhang2010, Crowdy2010,
  Crowdy2010b}.

In a simple model, we assume that the treadmilling organism has a
circular body of radius $\epsilon$, with a center at $z_d(t) = x_d(t)
+ i y_d(t)$.  We also assume that, with respect to the
angle~$\theta(t)$ of the head of the treadmilling organism from the
real axis, surface actuators of the treadmilling organism induce a
tangential velocity profile given by~\cite{Crowdy2010}
\begin{equation}
  U(\phi, \theta) = 2 V \sin(2(\phi-\theta)),
 \label{eq:U}
\end{equation}
where $V$ is a constant and $\phi$ is the angle measured from the
positive $x$ direction.

Next consider the treadmiller near an infinite wall along the~$x$
axis.  To satisfy the no-slip boundary condition at the wall, we
make the expansion
\begin{equation}
\begin{split}
  f(z, t) & = \frac{\mu}{z - z_d(t)} + f_0 + (z - z_d(t))f_1 + \cdots, \\
  g^\prime(z,t ) & = \frac{b}{(z - z_d(t))^3}
                      + \frac{a}{(z - z_d(t))^2} + g_0 + \cdots,
\label{eq:goursat}
 \end{split}
\end{equation}
$f(z, t)$ having no Stokeslet term and $g(z, t)$ having no rotlet term
implying that the treadmilling organism is force-free and
torque-free.  We use the boundary condition to find
\begin{equation}
  \mu = - i \epsilon \overline{c}, \qquad
  a = \mu {\bar z}_d, \qquad
  b = \mu \epsilon^2 - i \overline{c} \epsilon^3 = 2 \mu
     \epsilon^2,
 \label{eq:coeff_inside}
\end{equation}
where~$c(t) \equiv - i V \exp(-2i \theta(t))$.  We set the time scale
of the motion by letting $V = \epsilon^{-1}$ so that~$\mu(t) = \exp
(2 i \theta(t))$.  The coefficients~$f_0$, $f_1$, and $g_0$
are given in~\cite{Crowdy2010}.

The time derivative of the position and orientation of the
treadmilling organism is obtained by equating the time rate of change
of position to the local fluid velocity, and the time rate of change
of orientation to half the local vorticity:
\begin{equation}
 \frac{d z_d}{dt} = - f_0 + z_d {\bar f}_1 + {\bar g}_0,\qquad
 \frac{d \theta}{dt} = -2\, {\rm Im}\, f_1.
  \label{eq:governing_position_and_angle}
\end{equation}
Equation~\eqref{eq:governing_position_and_angle} can then be solved as
a set of three ODEs determining the motion of the treadmiller.

Now we turn to a semi-infinite wall, extending along the negative $x$
axis.  The conformal mapping~$\zeta = i z^{1/2}$ maps the~$z$ plane to
upper-half $\zeta$ plane, with the negative $x$-axis of the~$z$ plane
mapped to the real axis in the~$\zeta$ plane.  In the~$\zeta$ plane we
can use a similar singular expansion as~\eqref{eq:goursat}, but we
must take care to map the boundary conditions to the~$\zeta$ plane.
We omit the lengthy details, which are similar to Crowdy and
Samson~\cite{Crowdy2010b}.  See~\cite{ObuseGFD2010} for a more
complete derivation.  All that is required for simulating the swimmer
trajectories are the coefficients that appear
in~\eqref{eq:governing_position_and_angle}, which are
\begin{subequations}
  \begin{equation}
    f_0 = \frac{\mu}{4 z_d}
    - \frac{\epsilon^2 {\bar\mu}}{4 \zzd^3 {\bar z}_d^{3/2}}
    + \frac{\left(2 |z_d|^2 - 2 {\bar z}_d^2
        - 3\epsilon^2\right) {\bar\mu}}
    {8 \zzd^2{\bar z}_d^2}
    + \frac{\left(2 |z_d|^2 + 2{\bar z}_d^2
        - 3\epsilon^2\right){\bar\mu}}
    {8 \zzd{\bar z}_d^{5/2}} ,
    \label{f0}
\end{equation}
\begin{multline}
  f_1 = \frac{1}{12z_d^2}\biggl(-\frac{3 \mu }{4}
    + \frac{9 z_d^{3/2}\epsilon^2 {\bar\mu}}
    {2 \zzd^4 {\bar z}_d^{3/2}}
    - \frac{3z_d^{3/2} \left(2 |z_d|^2 - 2 {\bar z}_d^2
        - 3\epsilon^2\right) {\bar\mu}}
    {2 \zzd^3 {\bar z}_d^2} \\
    - \frac{3 z_d^{3/2} \left(2 |z_d|^2 + 2{\bar z}_d^2
        - 3\epsilon^2\right){\bar\mu}}
    {4 \zzd^2{\bar z}_d^{5/2}}\biggr),
  \label{f1}
\end{multline}
and
\begin{multline}
  g_0 = -\frac{3 \mu {\bar z}_d}{16 z_d^2}
  + \frac{10\epsilon^2 \mu}{32 z_d^3}
  + \frac{3 \epsilon^2{\bar\mu}}{8 \zzd^4 {\bar z}_d}
  - \frac{\left(z_d - {\bar z}_d\right){\bar\mu}}
  {4 \zzd^3 {\bar z}_d^{1/2}} \\
  + \frac{\left( 2|z_d|^2 - 6{\bar z}_d^2 - 3 \epsilon^2 \right)
    {\bar\mu}}
  {16 \zzd^2{\bar z}_d^{5/2}}
  \left({\bar z}_d^{1/2}+\zzd\right),
  \label{g0}
\end{multline}
\label{eq:f0f1g0}%
\end{subequations}
where $\zzd \equiv z_d^{1/2}+{\bar z}_d^{1/2}$.

\section{Results of numerical simulations}
\label{sec:NumericalResults}

We now present the results of numerical simulations of the governing
evolution equations~\eqref{eq:governing_position_and_angle}, together
with the coefficients~\eqref{eq:f0f1g0}.  The radius of the circular
treadmillers is set to~$\epsilon=1$, giving the reference length scale.
The time scale is set by~$V=\epsilon^{-1}$ in~\eqref{eq:U}.

\begin{figure}
 \centerline{\includegraphics[width=.8\textwidth]{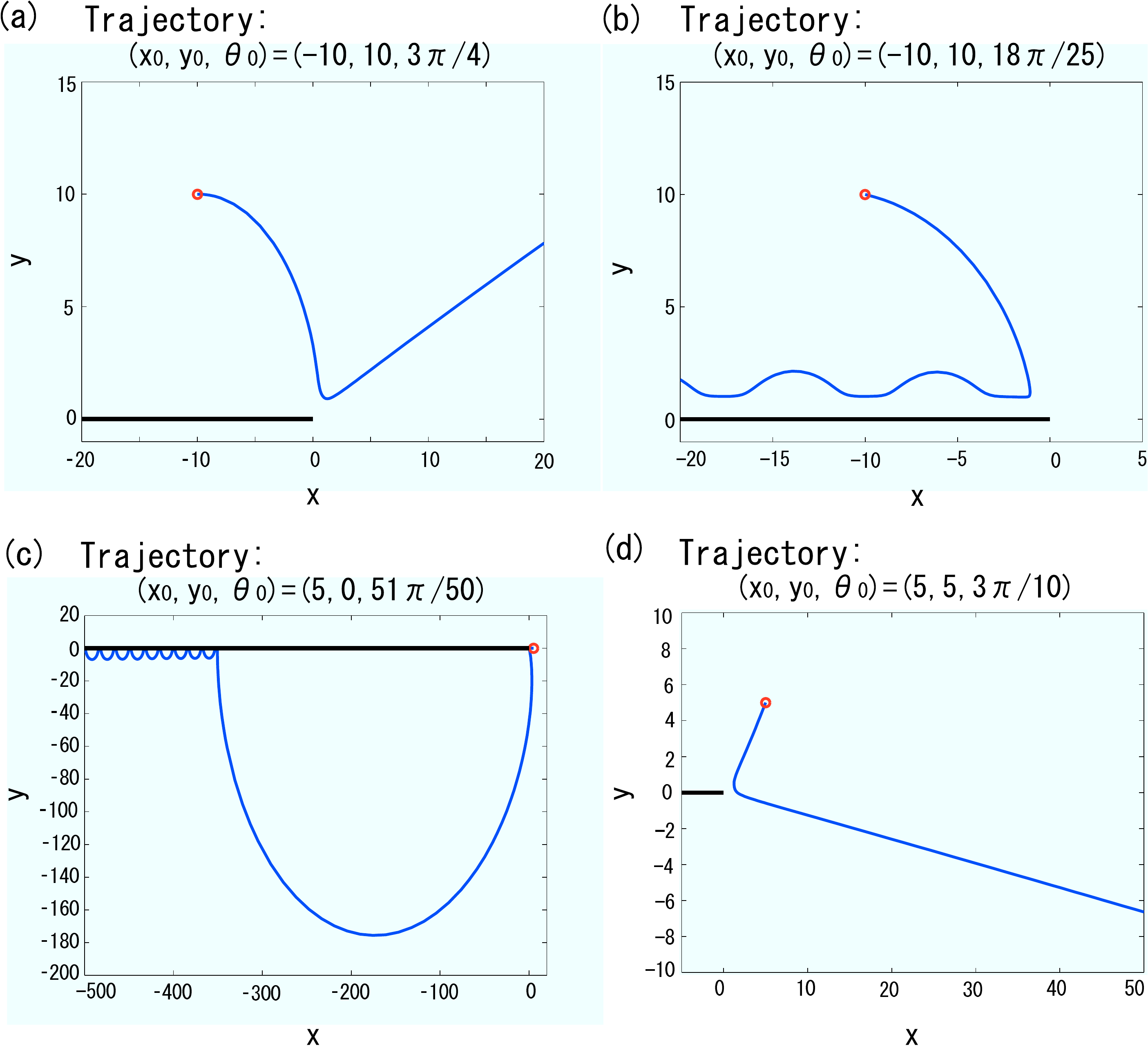}}
 \caption{Examples of trajectories from the initial points $(x_{d0},
   y_{d0})$ marked with circles, with an initial orientation
   $\theta_0$.  A treadmilling organism can escape from the wall, such
   as in (a) and (d), end up above the wall as in (b), or underneath
   the wall as in (c).}
 \label{fig:Examples}
\end{figure}

Figure \ref{fig:Examples} shows examples of swimmer trajectories for
different initial conditions for position~$z_{d0}$ and
angle~$\theta_0$.  Some trajectories, such as (a) and (d), end up far
away from the wall; we refer to those as escaping trajectories.
Others, such as (b) and (c), remain close to the wall for all time,
exhibiting the `bouncing' behavior noted in~\cite{Crowdy2010}.  A
fourth type of trajectory (not shown) crashes into the wall, but this
is an unphysical consequence of the boundary condition at the
swimmer's surface only being approximately satisfied.  It has been
verified numerically that the qualitative features of the trajectories
are not affected by this approximation~\cite{MattFinn_private}.

Both experimental observations and previous theoretical studies
suggest that, when there is a no-slip wall near a treadmilling
organism, the organism tends to be attracted to its own image and move
towards the wall \cite{Rothschild1963, Winet1984, Lauga2006,
  Berke2008, Ramia1993, HernandezOrtiz2005, Crowdy2010,
  Crowdy2010b}. This behavior is clearly seen at an early stage in all
the trajectories in Fig.~\ref{fig:Examples}. Nevertheless, in the cases
shown in Figs.~\ref{fig:Examples}(a) and (d), the treadmilling organism
moves away from the wall after it has come close to the edge of the
wall.

\begin{figure}
 \centerline{%
 \subfigure[]{%
   \includegraphics[width=.37\textwidth]{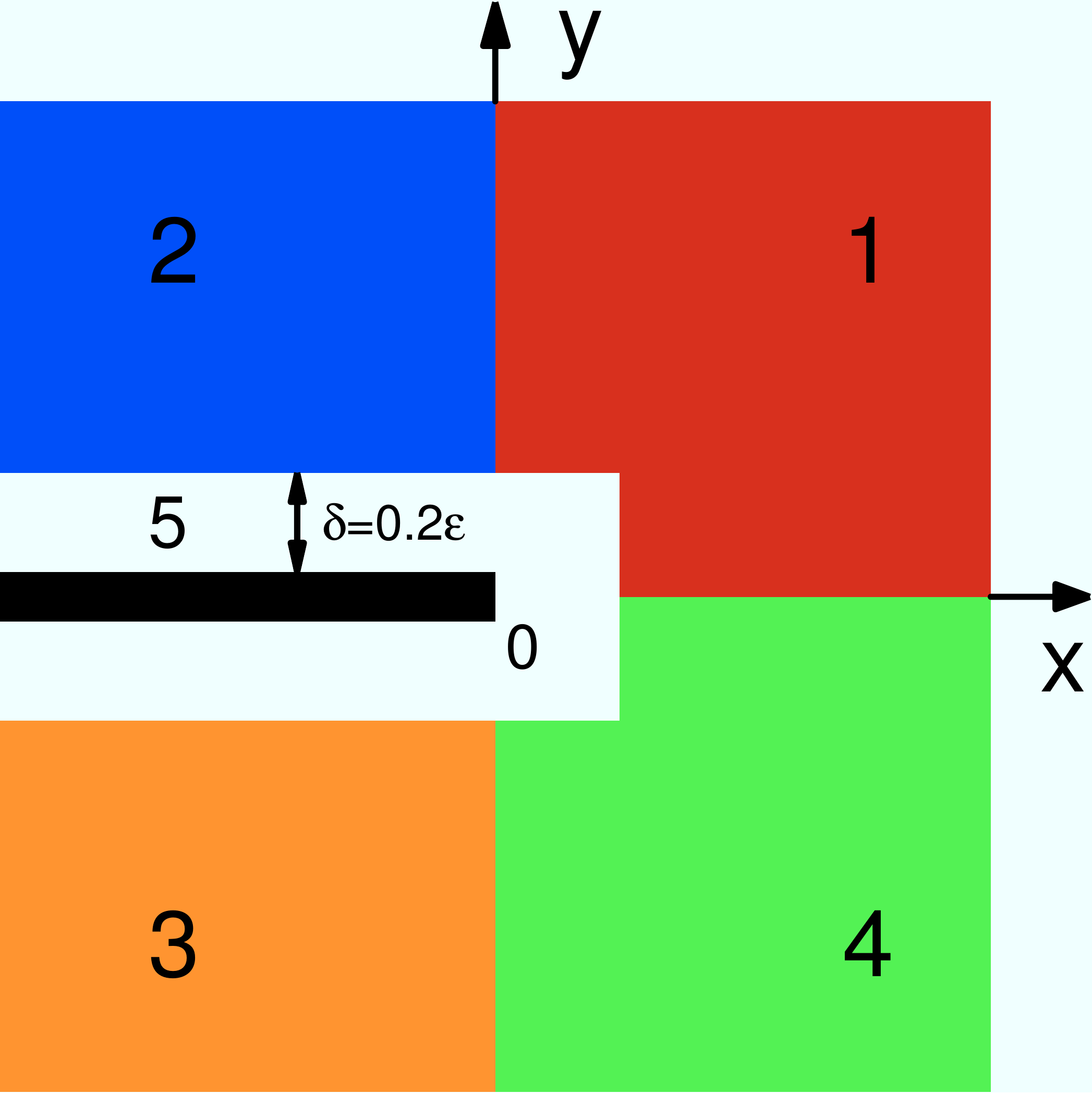}
 }\hspace{2em}%
 \subfigure[]{%
   \includegraphics[width=.37\textwidth]{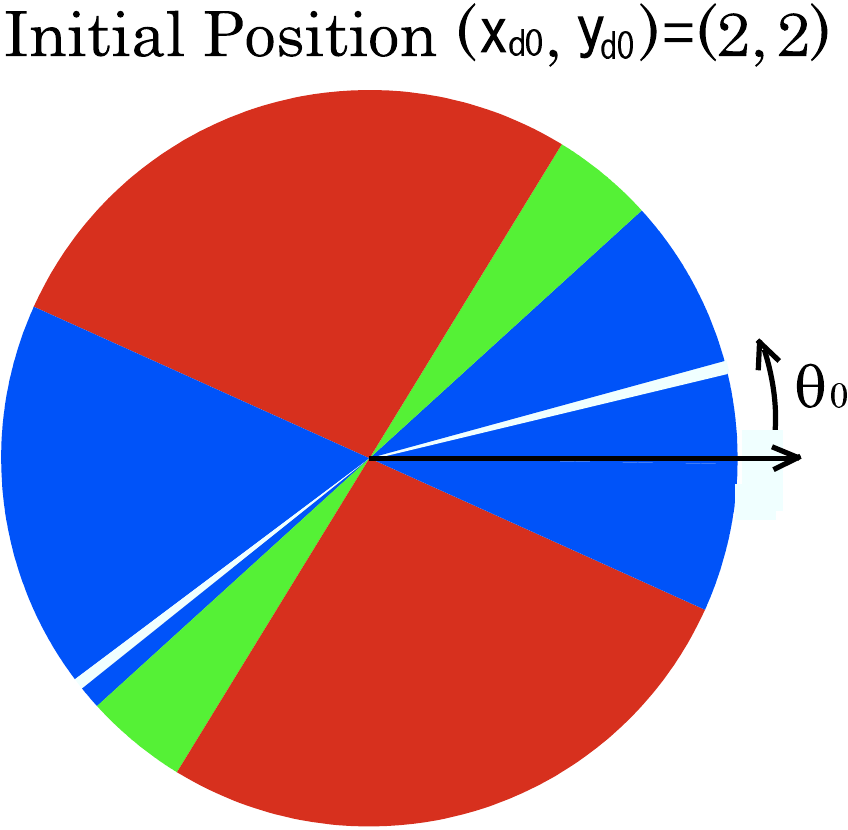}
 }}
\caption{(a) The~$x$--$y$ plane divided into five regions. The thick
  black line corresponds to the wall. (b) An example of a pie
  chart. The different \protect\ifcolor{colors}{shadings} as a
  function of initial orientation~$\theta_0$ indicate which region of
  (a) the treadmilling organism ends up in after a sufficiently large
  time.}
 \label{fig:FiveRegions}
\end{figure}

We now focus on the large-time behavior of treadmillers.  To study
this, we first introduce five regions shown in
Fig.~\ref{fig:FiveRegions}(a).  Regions 1--4 are the usual quadrants
of the plane, but with a neighborhood of size~$0.2\epsilon$ around the
wall removed; this removed neighborhood is region 5.
\ifcolor{%
  The regions 1--5 are colored red, blue, orange, green, and white,
  respectively.
}{%
  The regions 1--4 are rendered in different shades of gray, and
  region 5 is rendered in white.%
  \footnote{See
  \texttt{http://arXiv.org/abs/11XX.XXXX} for color figures.}
}%
From these regions, we can make a `pie chart' around each point in the
plane, an example of which is shown in Fig.~\ref{fig:FiveRegions}(b).
The sectors of the pie chart correspond to a range initial
angle~$\theta_0$; the \ifcolor{color}{shading} of the sector describes
which region the treadmiller ends up in for large times (here $t =
1500$).  Region 5 corresponds to `crashing' trajectories.  Thus, for a
given position, the treadmiller may end up in different regions, and
the relationship between angle and final region is not simple.

\begin{figure}
 \centerline{\includegraphics[width=\textwidth]{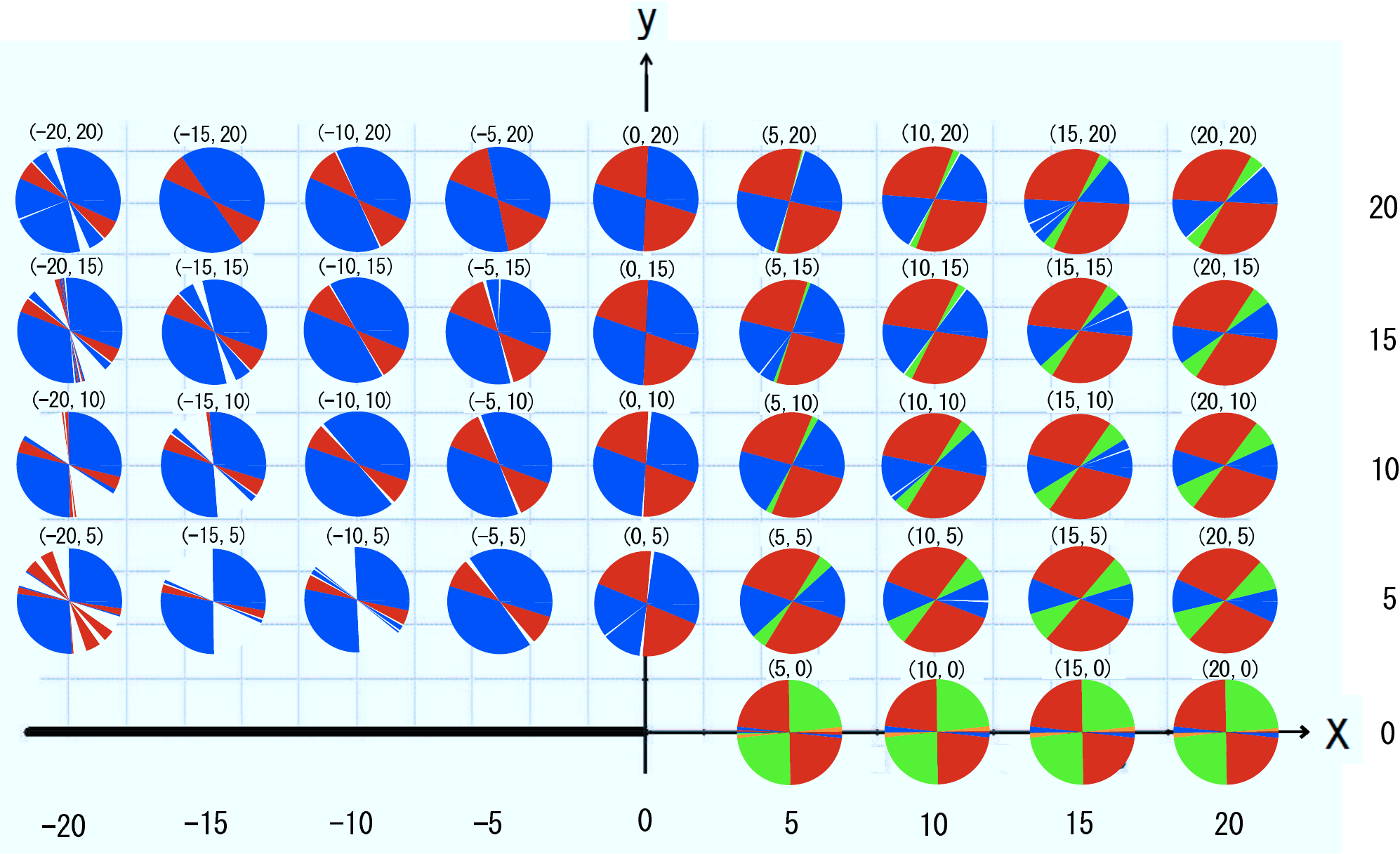}}
 \caption{Pie charts at $t=1500$ for several initial conditions. Each
   pie chart is centered on the initial condition it corresponds to.
   The thick black line represents the wall.  \protect\ifcolor{Note}{To help
     distinguish the shadings, note} that region 3 only appears as very
     thin slivers in the pie charts along the $x$ axis.}
 \label{fig:PieCharts}
\end{figure}

Figure \ref{fig:PieCharts} shows pie charts for many initial points
$(x_{d0}, y_{d0})$.  The most notable feature is the complexity of the
pie charts for initial points near the wall and with large negative
$x$ coordinates. This reflects the fact that the treadmilling organism
changes its heading direction and the quality of its trajectory
significantly when it has come to the vicinity of the edge of the
wall, $x = y = 0$ depending upon its $(x_d, y_d, \theta)$ at the time,
and so even a very small difference in initial condition can result in
a huge difference to its trajectory.  (This is an example of chaotic
scattering.)  However, pie charts tend to become rather
simple for larger $y_{d0}$ for any fixed $x_{d0}$, since the
treadmiller is then less influenced by the wall.

\begin{figure}
 \centerline{\includegraphics[width=.9\textwidth]{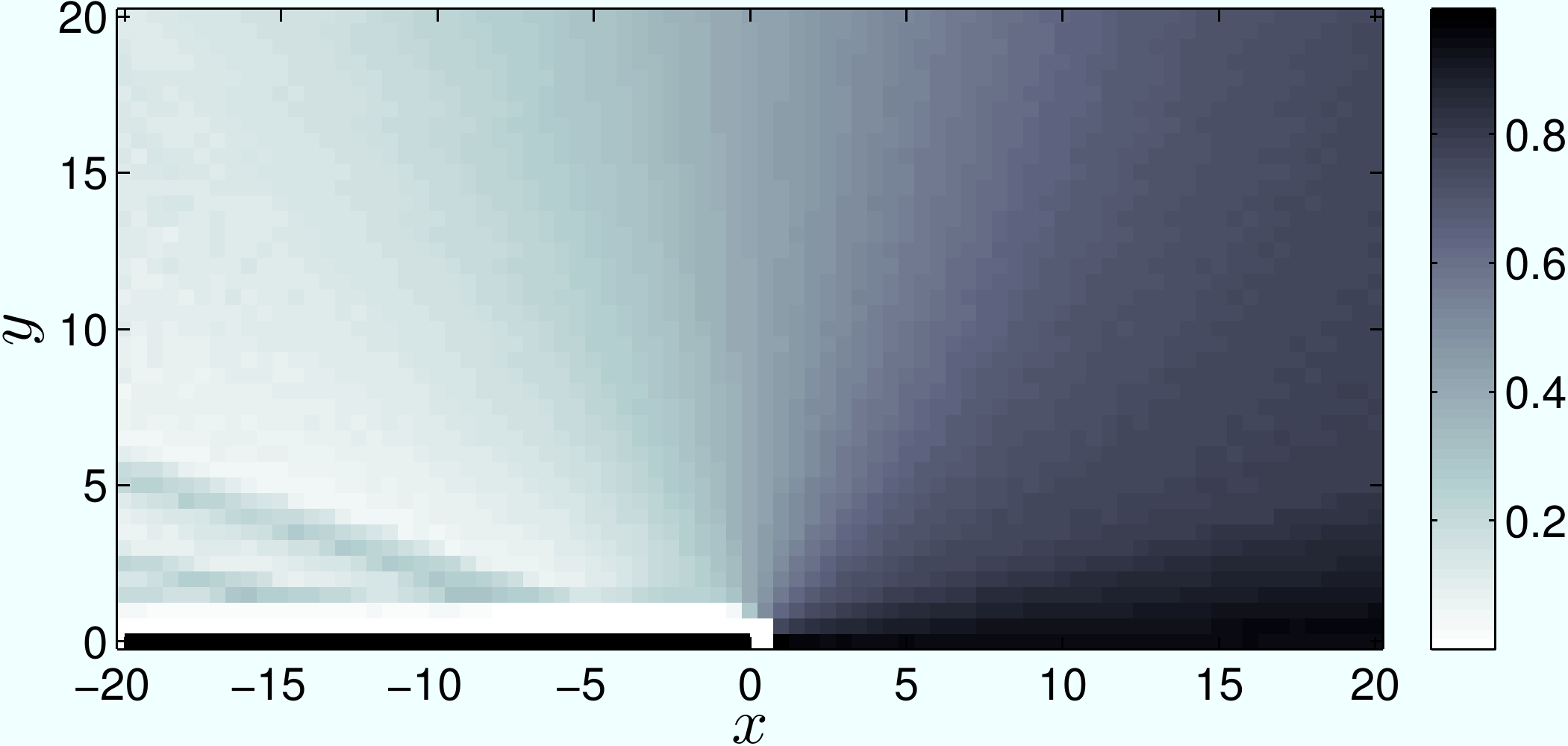}}
 \caption{Escape probability~$P_E$ at $t=1500$ for different initial
   points. Each point corresponds to a given initial
   condition~$(x_{d0},y_{d0})$, integrated over all possible starting
   angles~$\theta_0$.  A trajectory escapes if it ends up in regions 1
   or 4, as defined in Fig.~\ref{fig:FiveRegions}.  The excluded white
   area near the wall corresponds to initial conditions where the
   swimmer would be partially inside the wall.}
 \label{fig:EscapeProbability}
\end{figure}

Now let us consider the \emph{escape probability},
$P_E(x_{d0},y_{d0})$, the probability that the treadmilling organism
can escape from the wall region.  We define this as the probability
that the treadmilling organism be in region $1$ or $4$ at sufficiently
large time.  We assume that the initial angle~$\theta_0$ is uniformly
distributed.  The escape probability~$P_E(x_{d0},y_{d0})$ then
corresponds to the fraction of angles in a pie chart that \ifcolor{are
  red or green}{lead to region 1 or 4}, for a given initial
position~$(x_{d0},y_{d0})$.  The escape probability for $t = 1500$ for
different initial conditions is shown in
Fig.~\ref{fig:EscapeProbability}, where each $(x, y)$ coordinate
corresponds to an initial position of the treadmiller.  Note that
for almost all initial conditions there is some finite probability of
escaping or being trapped, though the escaping probability approaches
unity along the positive $x$ axis, and goes to zero for large negative
$x$ and large $y$.  Observe the strong `wedge' of trajectories that
are likely to escape ($P_E>0.6$), though there is also a
backwards-facing wedge of trajectories that have a reasonable change
of escaping ($P_E>0.3$).  This suggests that a probe or pipette
inserted in a medium is likely to `push away' treadmillers to some
degree.  Note also the `tongues' of abnormally-high ($P_E\simeq0.4$)
escape probability near the wall, which reflects the complicated
structure of the pie charts in that region.

\section{Discussion and conclusions}
\label{discussion_conclusion}

Following~\cite{Crowdy2010,Crowdy2010b}, we derived the governing
evolution equations for a treadmilling microorganism near a
semi-infinite no-slip wall.  This can also be regarded as a
two-dimensional model of a probe or pipette inserted in the fluid.  We
then numerically calculated the trajectories of the organism for
different initial conditions.  The treadmilling organism was usually
attracted to the wall for early times, but for later times often
escaped the wall region.  Typically this happened when the organism
came close to the tip of the wall.  To investigate this behavior
further, we looked at `pie chart' diagrams, where the sectors indicate
which region a swimmer eventually ends up in as a function of its
initial angular orientation.  Initial points with larger negative $x$
coordinates have more complex pie charts compared to those with
smaller negative $x$ coordinates, because they are sensitive to
`chaotic scattering' off the tip of the wall.

We then examined the escape probability $P_E$, the probability that
the treadmilling organism can escape to the right of the wall region,
assuming that it is initially randomly oriented.  There is an evident
`wedge' of initial conditions that is likely to escape the wall.  This
suggests that a probe or pipette inserted could `push' the organisms
out of the way (with the wedge replaced by a cone in three
dimensions), though since the organisms slow down as they get further
away from the wall the effect might not be very pronounced.

\section*{Acknowledgments}

The authors are grateful for the hospitality of the Geophysical Fluid
Dynamics Program at the Woods Hole Oceanographic Institution
(supported by NSF), and thank Matthew D.\ Finn for his helpful advice
and suggestions.  Some of the numerical calculations for this project
were performed at the Institute for Information Management and
Communication of Kyoto University.


\end{document}